# $n$-vicinities method for three dimensional Ising Model

**Boris Kryzhanovsky and Leonid Litinskii**

Scientific Research Institute for System Analysis of Russian Academy of Science, 117218 Nakhimovskii prospect Str. 36-1, Moscow, Russian Federation

E-mail: kryzhanov@mail.ru, litin@mail.ru

**Abstract**. The $n$-vicinities method for approximate calculations of the partition function of a spin system was proposed previously. The equation of state was obtained in the most general form. In the present publication these results are adapted to the Ising model on the $D$-dimensional cubic lattice. The state equation is solved for an arbitrary dimension $D$ and the behavior of the free energy is analyzed. For large values of $D$ ($D > 2$) the obtained results are in good agreement with the ones obtained by means of computer simulations. For small values of $D$, there are noticeable discrepancies with the exact results.

## 1. Introduction

In the papers [1]-[3] we develop the $n$-vicinities method for approximate calculation of the partition function. The method can be described as follows. Let $N >> 1$ be the dimension of the problem, and $\mathbf{s}_0 \in \mathbf{R}^N$ be the initial configuration whose $n$-vicinity $\Omega_n$ is the set of all the configurations that differ from $\mathbf{s}_0$ by the values of $n$ coordinates ($n = 0,1,...,N$). The distribution of the energies of the states from $\Omega_n$ fits the Gaussian density reasonable well [3]. The mean energy $E_n$ and the variance $\sigma_n^2$ can be accurately expressed in terms of the parameters of the connection matrix [1], [2]. Then in the asymptotic limit summing over the states from $\Omega_n$ in the expression for the partition function can be replaced by integration of $\exp(-\beta E)$ over the Gaussian measure, and summing over the all n-vicinities by integration between 0 and 1 over the parameter $x = n / N$. As a result the partition function takes the form of the integral $Z_N \approx \int dx \int dE \exp[-N \cdot F(x,E)]$, which is calculated with the aid of the saddle-point method. The function of two variables $F(x,E)$ depends on the following parameters: the inverse temperature $\beta$, the external magnetic field (if it is present), the trace of the connection matrix squared and some other numerical parameters. That is $F(x,E) \equiv F(x,E,\beta,H...)$. In the most general form the expression for $F(x,E)$ was obtained in [2], [3]. In the present publication we adapt this expression for the Ising model on the $D$-dimension hypercube, derive the equation of state, plot and examine the graphs of the free energy. Because of the size of the publication here is only the summary of our calculations. The details will be presented in [4].

## 2. Basic Expressions

Let $\mathbf{T}=\left(T_{ij}\right)_1^N$ be the connection matrix corresponding to the $D$-dimensional Ising model, $\mathbf{s}_0=(1,1,...,1)\in\mathbf{R}^N$ is the ground state of the spin system, $\Omega_x$ is the $x$-vicinity of the ground state, where $x=n/N\in[0,1]$. Let us calculate the energy per one spin: if $\mathbf{s}\in\Omega_x$, and $H$ is the value of the homogeneous magnetic field, then

$$E(\mathbf{s},H)=-\frac{1}{2N}\sum_{ij}T_{ij}s_is_j-\frac{1}{N}H\sum_i s_i=E(\mathbf{s})-H(1-2x).$$

By $E_0$ we denote the energy of the ground state: $E_0=E(\mathbf{s}_0)=-\frac{1}{2N}\sum_{ij}T_{ij}=-D$. Then the asymptotic expression for the energy averaged over $\Omega_x$ is $E_x=E_0(1-2x)^2$, and the variance is equal to $\sigma_x^2=\frac{8\sum_{ij}T_{ij}^2}{N^2}x^2(1-x)^2$. In [1]-[3] it was shown that the expression for the partition function can be reduced to the form: $Z_N\sim\int\limits_0^1 dx\int\limits_{E_0}^{|E_0|}\exp\left[-N\cdot F(x,E)\right]dE$, where $N>>1$,

$$F(x,E)=L(x)+\beta\left[E-H\cdot(1-2x)\right]+\frac{1}{2}\left(\frac{E-E_x}{\sigma_x}\right)^2,$$

and $L(x)=x\ln x+(1-x)\ln(1-x)$.

To find the global minimum of the function $F(x,E)$ it is necessary to solve the set of equations:

$$\frac{\partial F}{\partial E}=\beta+\frac{E-E_x}{\sigma_x^2}=0,\quad \frac{\partial F}{\partial x}=\ln\frac{x}{1-x}+\frac{E-E_x}{\sigma_x}\frac{\partial}{\partial x}\left(\frac{E-E_x}{\sigma_x}\right)+2\beta H=0.$$

Solving the first equation for $\beta$ and substituting the result into the second one, we obtain the minimization problem for the function of one variable $f(x)=L(x)+\beta E_x-\frac{\beta^2\sigma_x^2}{2}-\beta H(1-2x)$ restricted to the inequality $\beta+\frac{E_0-E_x}{\sigma_x^2}\le 0$. The restriction means that among all the minimums we examine only those for which $\frac{\partial F}{\partial E}=0$. We use the other normalization and more convenient notations:

$$b=\beta\frac{\sum_{ij}T_{ij}^2}{\sum_{ij}T_{ij}},\quad \gamma=\frac{\left(\sum_{ij}T_{ij}\right)^2}{N\sum_{ij}T_{ij}^2},\quad h=\frac{2NH}{\sum_{ij}T_{ij}}.$$

Then the final form of the problem is as follows: for each value of the "renormalized" inverse temperature $b$ we have to find the global minimum of the function

$$f(x)=L(x)-\frac{\gamma b}{2}\left[(1-2x)^2+8b(x(1-x))^2+h(1-2x)\right] \tag{1}$$

at the interval $(0,x_b)$, where

$$x_b=\begin{cases}\frac{1}{2}, & \text{when } b\le 1\\ \frac{1-\sqrt{1-1/b}}{2}, & \text{when } b\ge 1\end{cases}. \tag{2}$$

When $b$ exceeds 1, the right-hand boundary of the interval where we are looking for the global minimum becomes depending on $b$.

The free energy is equal to $f(b)=\min\limits_{x\in(0,x_b)}f(x)$.

**Note.** For the $D$-dimensional Ising model with the connection $J$ between the nearest neighbors we have: $\sum_{ij} T_{ij} = DNJ$, $\sum_{ij} T_{ij}^2 = DNJ^2/2$, $b = \beta J$, $\gamma = 2D$, $h = H/(DJ)$. Here the characteristic $\gamma$ is equal to the number of the nearest neighbors of each spin. However, generally the interactions along the different directions of the lattice can be different. In this case $\gamma$ is no longer an integer. For example, for the two-dimensional Ising model with different interaction constants along the horizontal ($J$) and vertical directions ($K$) we have $\gamma = 2\left(1 + \frac{2}{K/J + J/K}\right)$. Then $\gamma$ can takes any value from the interval $2 < \gamma < 4$. Similarly, for the three-dimensional Ising model with different interaction constants ($J, K, L$), $\gamma$ can takes any value from the interval $2 < \gamma < 6$. In the general case, $\gamma$ is the effective coordination number that describes the interaction of the spin with the nearest vicinity.

### 3. Analysis of the state equation when *H*=0

Let us set $H = 0$. The case of the nonzero magnetic field will be described in [4]. We seek the global minimum of the function $f(x)$ (1) with $h = 0$. For that we have to solve the state equation

$$\frac{\partial f(x)}{\partial x} = \ln\frac{x}{1-x} + 2\gamma b(1-2x)\left[1 - 4bx(1-x)\right] = 0. \quad (3)$$

It is evident that $x_0 = 0.5$ is the solution of this equation for any values of $b$ and $\gamma$. $x_0$ is called the trivial solution. For very small values of $b$ ($b \sim 0$) the function $f(x)$ decreases monotonically from $f(0) = -\gamma b/2$ up to $f(x_0) = -\ln 2 - \gamma b^2/4$ (see figure 1a). Consequently, for small $b$ the trivial solution $x_0$ is the only solution of equation (3), and the free energy takes the form

$$f(b) = \min_x f(x) = -\ln 2 - \frac{\gamma b^2}{4}, \quad b \sim 0. \quad (4)$$

**The interval** $\gamma > 16/3$. It can be shown [4] that for $\gamma$ from this interval the point $x_0$ remains the only solution of equation (3) until $b$ is less than the critical value

$$b_c = \frac{1 - \sqrt{1 - 4/\gamma}}{2}. \quad (5)$$

When $b < b_c$, the free energy has the form (4). Once $b$ exceeds the value $b_c$, $x_0$ becomes the maximum point of the function $f(x)$. To the left from $x_0$ appears another minimum point $x_1(b)$: $x_1(b) < x_0$ (see figure 1b). In this case, to calculate the free energy one has to substitute $x_1(b)$ in equation (1). When $b = b_c$ the phase transition of the second kind takes place in the system.

A further increase of the parameter $b$ is accompanied by the deepening of the minimum and steady shifting of the minimum point towards 0. When $b$ becomes larger than 1, the right boundary of the interval becomes the variable that is equal to $x_b$ (see equation (2)). When $b \to \infty$ the interval $(0, x_b)$ contracts to the origin of coordinates and the global minimum $x_1(b)$ tends to zero. As it has to be, when the temperature tends to zero, the spin system tends to its ground state $\mathbf{s}_0 = (1, 1, ..., 1)$.

**Table 1.** Theoretical and experimental values of $b_c$ for D-dimensional Ising models

| | 3D | 4D | 5D | 6D | 7D |
|---|---|---|---|---|---|
| Our theory | 0.2113 | 0.1464 | 0.1127 | 0.0918 | 0.0774 |
| Computer simulations | 0.2216 | 0.1489 | 0.1139 | 0.0923 | 0.0777 |

Let us substitute the values $\gamma = 6, 8, 10, 12, 14$ into equation (5) and calculate the critical values of the inverse temperature for the 3D-, 4D-, 5D-, 6D- and 7D- Ising models, respectively. The obtained numbers are shown in the upper row of the table 1. In the lower row we present the values of $b_c$ that are the results of computer simulations (see [5]-[7]). We see a good agreement between our theoretical estimates and the results of computer simulations.

Thus, equation (5) provides us with a reasonable estimate of the critical temperature for the Ising model on the D-dimension hypercube when $D > 8/3$. We would like to remind that the parameter $\gamma$ has not to be even integer-valued numbers (see the note in the end of the previous section).

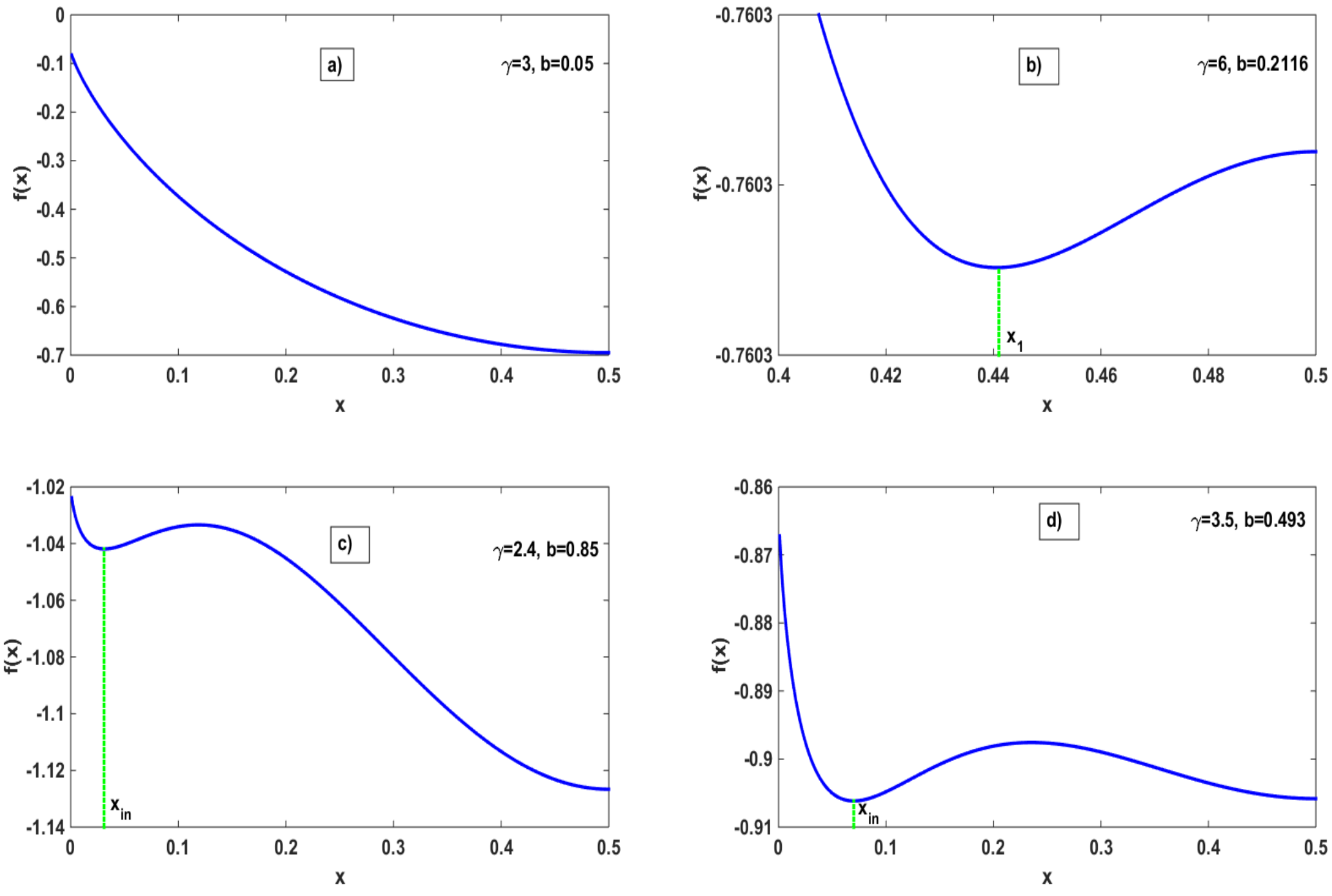

**Figure 1.** The behaviour of the function $f(x)$ for different values of $\gamma$ and $b$ (see the body of the text).

When $\gamma < 16/3$, with increase of the parameter $b$ the behavior of the function $f(x)$ changes. Namely, now it certainly has a minimum in the inner point $x_{in}(b)$ of the interval $(0, x_0)$. The depth of this minimum has to be compared with $f(x_0)$. Different situations are possible depending on whether $\gamma$ is less or larger than 2.75.

**The interval** $\gamma < 2.75$ (figure 1c)**.** It turns out that in this case the local minimum in the point $x_{in}$ is never deeper than the minimum in the point $x_0$ [4]. Then as long as $b \le 1$ the global minimum of the function $f(x)$ is in the point $x_0$. When the value of $b$ becomes larger than 1 and the right boundary $x_b$ becomes depending on $b$, and the global minimum of the function $f(x)$ is in the point $x_b$:

$$f(b) = \begin{cases} -\ln 2 - \dfrac{\gamma b^2}{4}, & b \le 1 \\ L(x_b) - \dfrac{\gamma}{4}(2b-1), & b \ge 1 \end{cases}. \qquad (6)$$

During the further increase of $b$ ($b \to \infty$) the global minimum is either in the point $x_b$ or it jumps to the point of the current minimum $x_{in}(b)$. The last happens if for a value of $b$ the minimum $f(x_b)$ becomes equal to $f(x_{in})$. In any case, when $b$ grows indefinitely, the minimum point tends to 0 and the system tends to the ground state.

Note, when $b=1$ there is a discontinuous of the first derivative of the free energy (6): $f_b'|_{b=1-0} = -\gamma/2$ and $f_b'|_{b=1+0} = (1-\gamma)/2$, which demonstrates the presence of the phase transition of the first kind. This is true for any $\gamma < 2.75$, in particular for $\gamma = 2$ that corresponds to the 1D Ising model. In the same time, the exact solution for the one-dimension Ising model does not show phase transitions for finite temperatures [8]. We have to admit that in the region $\gamma < 2.75$ our method provides questionable results.

**The interval** $2.75 < \gamma < 16/3$ (figure 1d). In this case, for a value of $b$ the inner local minimum $x_{in}(b)$ becomes necessarily equal to the minimum in the point $x_0$ [4]. It takes place when $b = b_j$ and by $x_j = x_{in}(b_j)$ we denote the corresponding value of $x$. For $b = b_j$ the global minimum jumps from the point $x_0$ to the point $x_j$. The magnetization of the system changes in a step-wise way from $m_0 = 0$ to $m_j = 1 - 2x_j$, indicating the phase transition of the first kind. With further increase of $b$ ($b \to \infty$) the minimum point $x_{in}(b)$ tends to 0 steadily and the system tends to its ground state.

The value $\gamma = 4$ that corresponds to the 2D Ising mode belongs to the interval in question: $\gamma \in [2.75, 16/3]$. From our approach it follows that in the 2D Ising system the phase transition of the first kind occurs when $b_c \approx 0.3912$. In the same time, the exact solution of Onsager shows the phase transition of the second kind for the noticeably larger value $b_c \approx 0.4407$ [8]. When $2.75 < \gamma < 16/3$ the results of our method are also questionable.

## 4. Discussion and conclusions

Let us illustrate the above statements by some computer calculations. In figure 2 for different values of $\gamma$ the dependences of the magnetization $m = 1 - 2x$ on the renormalized inverse temperature $b$ are shown. At first when $b$ is not too large, the global minimum is in the point $x_0 = 0.5$, and the magnetization of the state is equal to zero. When the parameter $b$ reaches the critical value the global minimum leaves the point $x_0$ and from this moment the magnetization is a nonzero one. For $\gamma = 6$ it happens when $b = b_c$ and after that the magnetization varies smoothly, without discontinuities. For other values of $\gamma$ it happens when $b = b_j$ and at that the global minimum jumps from the point $x_0$ into the point $x_j$ (see above). For $\gamma < 16/3$ the jump of the magnetization $m = 1 - 2x_j$ is clearly seen in figure 2.

The value of the mean energy $<E> = \sum_{\mathbf{s}} E(\mathbf{s}) \exp(-\beta \cdot E(\mathbf{s}))$ behaves itself in the same way. The cumbersome calculations of this characteristic will be also presented in [4]. In Fig.3 for different values of $\gamma$ the dependences of the renormalized mean energy $E = <E>/E_0$ as functions of $b$ are shown. At first $E$ is linear in $b$; when $b$ becomes larger than the critical value, the behavior of $E$ changes. When $\gamma = 6$, the function $E(b)$ is a continuous function that decreases up to $-1$. When $\gamma \in [2.75, 16/3]$ the value of $E$ has a jump in the critical point. After that the system tends to the ground state. The jump manifests the presence of the gap in the distribution of the energies of states. However, we have no arguments why there is the energy gap when $\gamma < 16/3$, and it is absent when $\gamma = 6$.

The essential fault of our approach is the presence of the jump of the magnetization in the 2D Ising model. This error is the result of distortions due to the approximation of the true distribution of energies by the Gaussian density. Without getting into details, let us note that analyzing the equation of state we see how a small distortion of the curve's shape allows to eliminate the unnecessary jump of the magnetization. We hope that this can be done by an improvement of our model. For example, a more accurate approximation can be used in place of the Gaussian density. This is the subject of our further analysis. Even in the present form when we ignore the jumps of the magnetization and are interested in the behavior of the free energy only, the results obtained with the aid of our approach are reasonable enough.

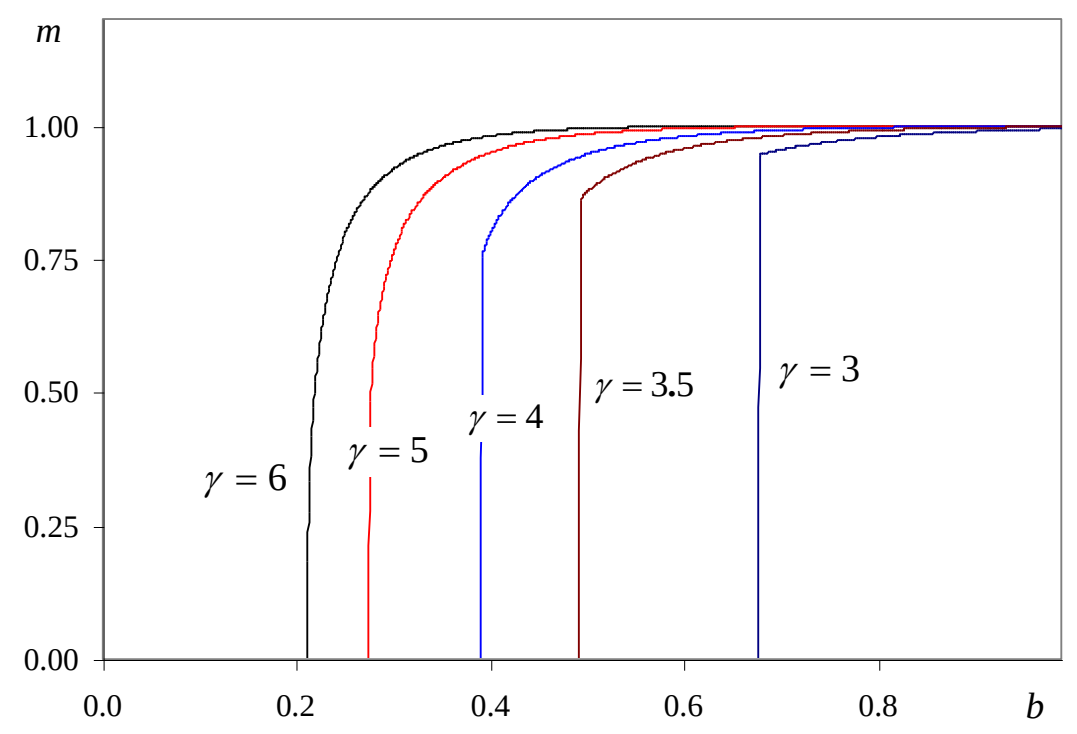

**Figure 2.** Magnetization $m$ as function of $b$ for different values of $\gamma$ .

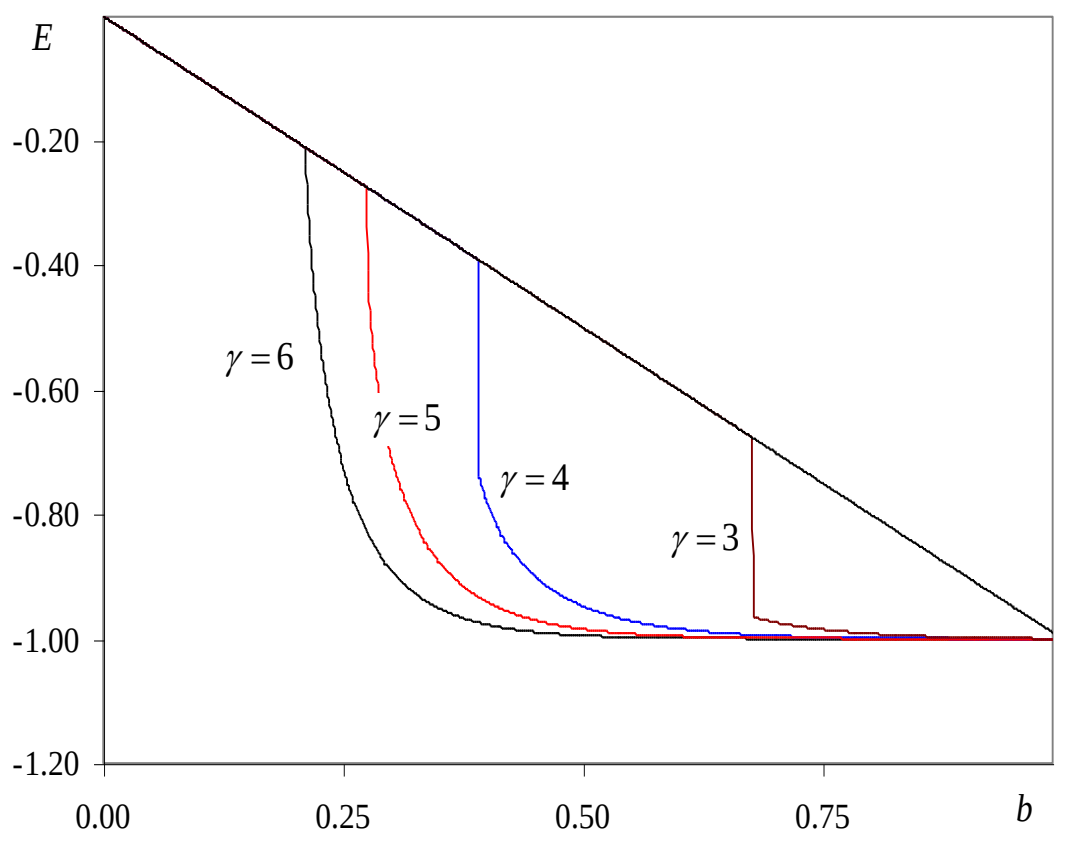

**Figure 3.** Renormalized mean energy $E$ as function of $b$ for different values of $\gamma$ .

In figure 4 for the 2D Ising model we show the dependence of the inverse critical temperature $\beta_c$ on the ratio of the interaction constants along the two axes: $J = 1$ and $K$ varies from 0.2 to 1. We see that our results are in sufficiently good agreement with the exact solution.

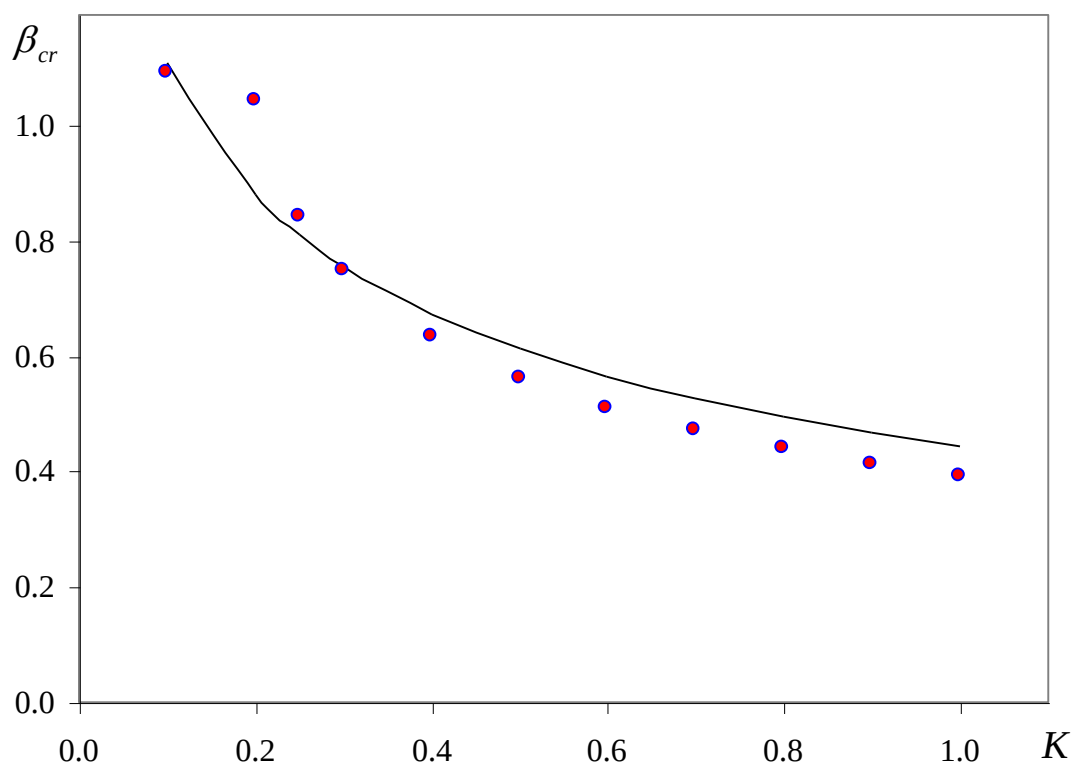

**Figure 4.** 2D Ising model ( $\gamma = 4$ ). Dependence of the critical temperature on K: solid line is the Onsager theory, circles is our results.

In figure 5 the graphs of the free energy for different values of $K$ are shown. The solid line is the free energy $f_{ex}(\beta)$ calculated according the Onsager theory, the circles are the results of our approach $f_{app}(\beta)$. It is clearly seen that the difference between our results and the exact theory are noticeable only in the region near the critical temperature $\beta_c$. To make this statement more evident, in figure 6 we present the graphs of the relative error $\left( f_{app}(\beta) - f_{ex}(\beta) \right) / f_{ex}(\beta)$. Its maximal value corresponds to the critical temperature $\beta_c$ and it is of the order of 1%.

Let us summarize. For cubic lattices of high dimensions ($\gamma > 16/3$) our theory is in good agreement with the well-known results. There are no exact results for the high dimensional Ising models ($D > 2$). For these dimensions our approach, as well as the computer simulations, predicts the phase transition of the second kind. The experimental values of the critical temperatures for a rather long sequence of the values of $\gamma$ coincide sufficiently accurately with the ones obtained with the aid of our formula (5) – see table 1.

When $\gamma < 16/3$ our theory is not sufficiently accurate. The most important fault is the presence of the phase transition of the first kind in the cases where there are no phase transitions at all ($\gamma = 2$), or the phase transition has to be of the second kind ($\gamma = 4$). The aim of our further analysis is to correct this defect.

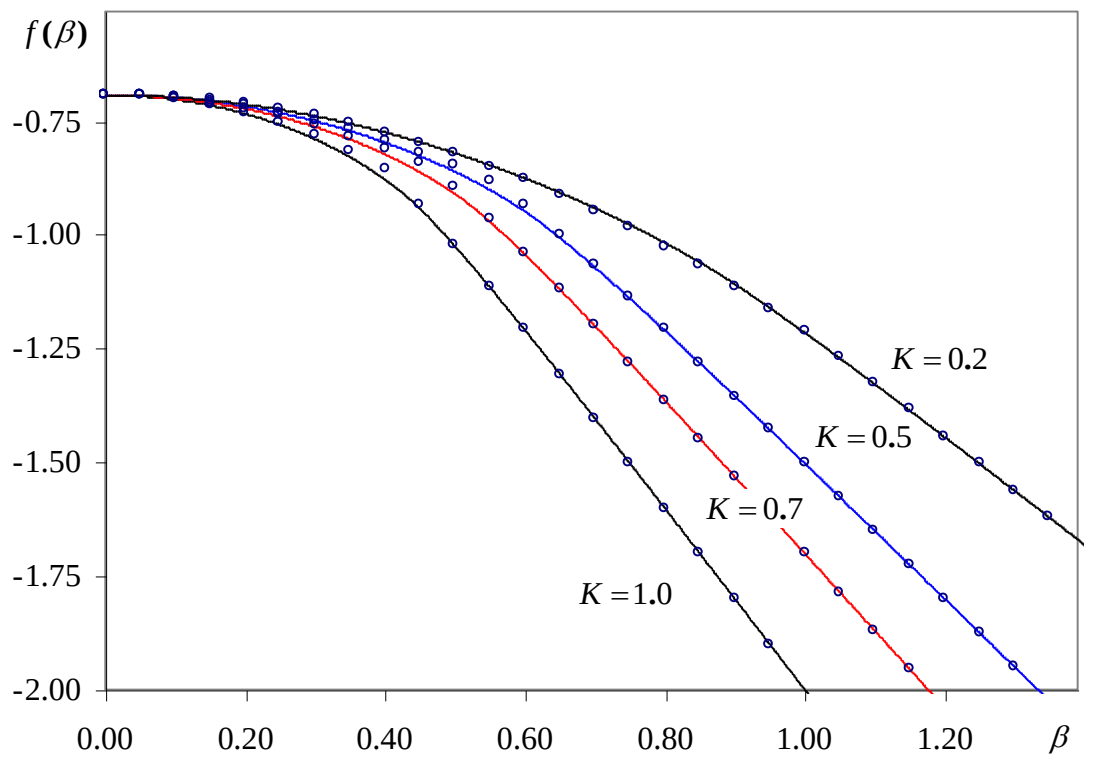

**Figure 5.** 2D Ising model ($\gamma = 4$). Free energy $f(\beta)$ for $K = 0.2,\ 0.5,\ 0.7,\ 1.0$: solid lines are the Onsager theory, circles are our results.

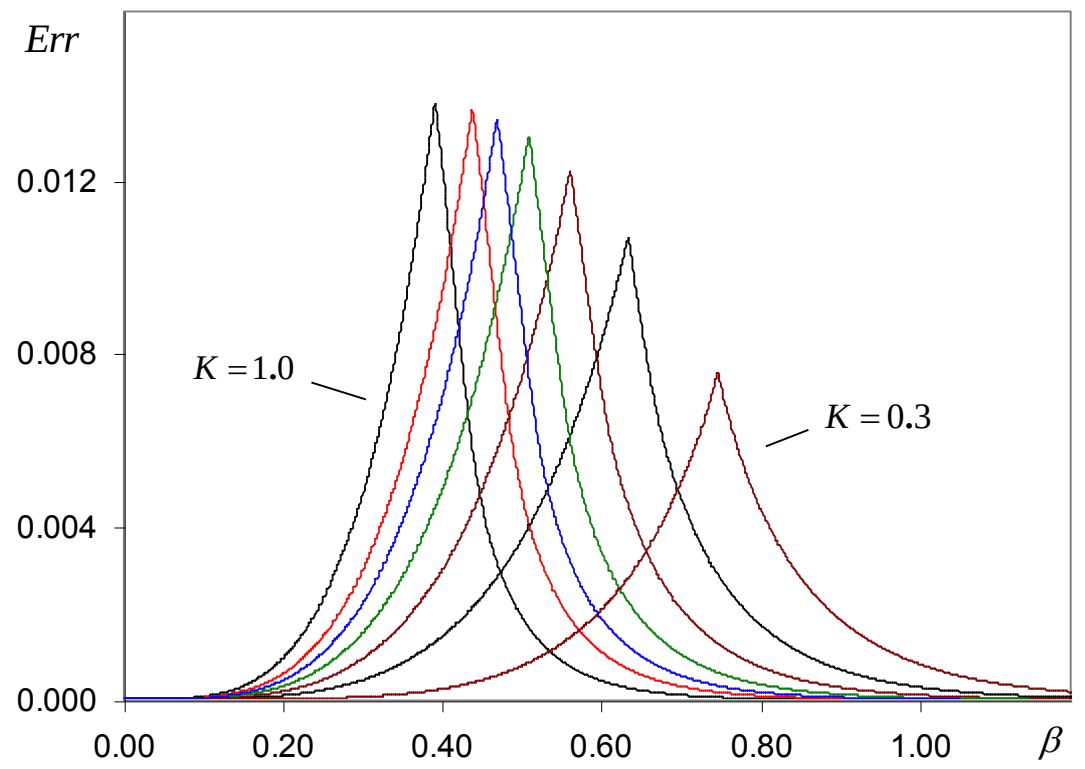

**Figure 6.** 2D Ising model ($\gamma = 4$). Relative errors as functions of $\beta$ for $K = 1, 0.8, 0.7, 0.6, 0.5, 0.4, 0.3$ (from left to right).

The work was supported by the financial support of RBRF grants №15-07-04861 and 16-01-00626.

**References**

[1] Kryzhanovskiy B, Litinskii L 2014 http://iopscience.iop.org/1742-6596/574/1/012017/pdf/1742-6596_574_1_012017.pdf

[2] Kryzhanovsky B V, Litinskii L B 2014 *Doklady Mathematics* **90** 1

[3] Kryzhanovsky B, Litinskii L 2015 *Opt. Mem. & Neu. Net.* (*Inf. Optics*) **24** 165

[4] Kryzhanovsky B, Litinskii L 2016 (in preparation)

[5] Blote H W J, Shchur L N, Talapov A L 1999 *Int. J. Mod. Phys.* C **10** 1137

[6] Lundow P H, Markstrom K 2004 The critical behaviour of the Ising model on the 4-dimensional lattice Preprint arXiv:1202.3031v1

[7] Lundow P H, Markstrom K 2015 *Nucl. Phys.* **B 895** 30

[8] Baxter R J 1982 *Exactly solved models in statistical mechanics* (London: Academic Press)